\documentclass[conference]{IEEEtran}
\IEEEoverridecommandlockouts
\usepackage{cite}
\usepackage{amsmath,amssymb,amsfonts}
\usepackage{algorithmic}
\usepackage{graphicx}
\graphicspath{{Fig/}}
\usepackage{textcomp}
\usepackage{xcolor}
\usepackage{tabularx}
\usepackage{subfigure} 
\usepackage{bm}
\usepackage{booktabs}
\usepackage{multirow}
\usepackage{makecell}
\usepackage{subfigure}

\usepackage[hidelinks]{hyperref}

\def\BibTeX{{\rm B\kern-.05em{\sc i\kern-.025em b}\kern-.08em
		T\kern-.1667em\lower.7ex\hbox{E}\kern-.125emX}}

\begin{document}

\title{Towards Next-Generation Intelligent Maintenance: Collaborative Fusion of Large and Small Models}

\author{
	\IEEEauthorblockN{Xiaoyi Yuan}
	\IEEEauthorblockA{\textit{Shenzhen ZNV Technology Co.,Ltd.} \\
		Shenzhen, China \\
		yuanxiaoyi@znv.com}
	\and
	\IEEEauthorblockN{Qiming Huang}
	\IEEEauthorblockA{\textit{Shenzhen ZNV Technology Co.,Ltd.} \\
		Shenzhen, China \\
		huangqiming@znv.com}
	\and
		\IEEEauthorblockN{Mingqing Guo}
	\IEEEauthorblockA{\textit{Shenzhen ZNV Technology Co.,Ltd.} \\
		Shenzhen, China \\
		guomingqing@znv.com}
	\and
		\IEEEauthorblockN{Huiming Ma}
	\IEEEauthorblockA{\textit{Shenzhen ZNV Technology Co.,Ltd.} \\
		Shenzhen, China \\
		mahuiming@znv.com}
	\and
		\IEEEauthorblockN{Ming Xu}
	\IEEEauthorblockA{\textit{Department of Automation} \\
		\textit{Tsinghua University}\\
		Beijing, China \\
		m-xu18@mails.tsinghua.edu.cn}
	\and
		\IEEEauthorblockN{Zeyi Liu}
	\IEEEauthorblockA{\textit{Department of Automation} \\
		\textit{Tsinghua University}\\
		Beijing, China \\
		liuzy21@mails.tsinghua.edu.cn}
	\and
		\IEEEauthorblockN{Xiao He}
	\IEEEauthorblockA{\textit{Department of Automation} \\
		\textit{Tsinghua University}\\
		Beijing, China \\
		hexiao@tsinghua.edu.cn}

	\thanks{This work was supported by National Natural Science Foundation of China under grants 62473223, 624B2087 and 62163012, and Beijing Natural Science Foundation under grant L241016. ({\it Corresponding author: Xiao He.})
	}
}

\maketitle
\begin{abstract}
With the rapid advancement of intelligent technologies, collaborative frameworks integrating large and small models have emerged as a promising approach for enhancing industrial maintenance. However, several challenges persist, including limited domain adaptability, insufficient real-time performance and reliability, high integration complexity, and difficulties in knowledge representation and fusion. To address these issues, an intelligent maintenance framework for industrial scenarios is proposed. This framework adopts a five-layer architecture and integrates the precise computational capabilities of domain-specific small models with the cognitive reasoning, knowledge integration, and interactive functionalities of large language models. The objective is to achieve more accurate, intelligent, and efficient maintenance in industrial applications. Two realistic implementations, involving the maintenance of telecommunication equipment rooms and the intelligent servicing of energy storage power stations, demonstrate that the framework significantly enhances maintenance efficiency.
\end{abstract}

\begin{IEEEkeywords}
Intelligent maintenance, Large Models, Small Models, Fusion
\end{IEEEkeywords}


\section{Introduction}

Intelligent maintenance plays a crucial role in enhancing system reliability and enabling timely decision-making in complex industrial environments \cite{11007216}. 
The rapid evolution of large language models (LLMs) offers new solutions to address the demands of intelligent maintenance in industrial scenarios. LLMs have been extensively applied to fault diagnosis and intelligent maintenance across various domains, achieving notable progress. In areas such as power grid fault diagnosis, and equipment vibration analysis \cite{liu2024fault,khan2024faultexplainer,qaid2024fdllm,dave2024integrating,tao2024llmbearing}, researchers have proposed novel approaches that integrate LLMs to overcome the limitations of traditional methods, thereby enhancing diagnostic accuracy and interpretability. In the field of intelligent maintenance, which encompasses sectors such as electricity \cite{he2024maintagt}, ports \cite{pei2024ai}, and railways \cite{chen2025ai}, the construction of dedicated models and innovative architectures has advanced the transition toward intelligent maintenance systems. For example, the MaintAGT professional large model \cite{he2024maintagt} and the LLM-based maintenance assistant for port equipment have significantly improved operational efficiency. In addition, to address challenges in mechanical equipment health management and Prognostics and Health Management (PHM), the PHM-LM framework has been proposed \cite{tao2024phmoutline}, providing novel development directions and technical pathways for realizing the intelligent transformation of industrial operations.

However, the above-mentioned work mainly relies on the traditional text understanding and reasoning capabilities of large models. In contrast, traditional small models (SMs) have inherent advantages in handling specific domains, structured data, and performing precise calculation tasks. Combining the advantages of both and constructing a technical framework for the fusion and collaboration of large and small models will be a crucial approach to improving the performance of intelligent maintenance systems in industrial scenarios.

This paper further reviews existing paradigms for the fusion and collaboration of large and small models. It presents a detailed technical framework for such integration, specifically designed for intelligent maintenance in industrial scenarios. The rationality and advancement of the proposed framework are analyzed through case studies involving the maintenance of telecommunication equipment rooms and the intelligent upkeep of new energy plants.

\section{Overall Framework}

\subsection{Basic Definitions}

Generally, a large model refers to a type of deep neural network model with a large number of parameters (with a scale of billions or more) \cite{min2024synergetic,zhang2024cogenesis}. Through pre-training on a large-scale dataset, it acquires the ability to comprehensively understand or generate data such as language and images. With this model, it can be adapted to specific downstream tasks without or with only a small amount of fine-tuning. According to the different data it processes, typical large models can be further divided into large language models and large multi-modal models.

Correspondingly, a small model refers to a model with a small amount of parameters (with a scale of one billion or less). If divided according to the core differences in the technical framework, small models can be further classified into two categories: SLM (Small Large Model) and traditional small models (business small models, vertical small models). Among them, the SLMs is a lightweight version obtained after distillation and quantization of the LLMs, and is applied to specific fields, such as deepseek-R1-1.5B; traditional small models refer to models with a small amount of parameters generated by traditional machine learning or deep learning frameworks such as CNN/RNN, such as CV models like Yolo/Resnet, or data models such as linear regression, decision tree regression, and Bayesian regression.
A brief comparison between large and small models is shown in Table~\ref{tab:model_comparison}.

\begin{table*}[htbp]
	\centering
	\caption{Comparison Between Large and  Small Models}
	\renewcommand{\arraystretch}{1.3}
	\setlength{\tabcolsep}{3mm}{
		\begin{tabular}{{p{3.2cm}p{3.7cm}p{3.7cm}p{4.9cm}}}
			\specialrule{0.1em}{1pt}{1pt}
			\specialrule{0.1em}{1pt}{3pt}
			\textbf{Dimension} & \textbf{LLM (Large)} & \textbf{LLM (Small)} & \textbf{Traditional small models} \\
			\midrule
			Number of parameters & Trillion-level ($>$100B) & Billion-level ($\sim$1B) & Million to sub-billion (1M--0.1B) \\
			Hardware requirements & GPU/TPU clusters & Single GPU & Single GPU, edge devices \\
			Latency & High (second-level response) & Moderate & Low (millisecond-level response) \\
			Memory / Storage usage & Extremely high & Moderate & Rather low \\
			Applicable Scenarios & Complex reasoning, multi-modal generation & Lightweight NLP (e.g., chatbot, text classification) & Real-time tasks, edge-side inference \\
			Cost & High training/inference cost & Moderate training/inference cost & Low training/inference cost \\
			Advantages & Strong generalization and multitask capability & Balanced performance and efficiency & High computational efficiency, fast training/inference \\
			Limitations & High resource consumption and deployment cost & Inferior reasoning capability compared to large models & Limited language understanding and generation \\
			Typical models & Deepseek-V3-671B & Deepseek-r1-1.5B & XGBoost, Yolo-v8 \\
\specialrule{0.1em}{3pt}{1pt}
\specialrule{0.1em}{1pt}{1pt}
	\end{tabular}}
	\label{tab:model_comparison}
\end{table*}

\subsection{Collaboration Paradigm}

In recent years,  various frameworks have been proposed  for the fusion and collaboration of large and small models, aiming to utilize the general knowledge and understanding capabilities of LLMs as well as the domain expertise and computational accuracy of Small Models (SMs). These frameworks can be roughly summarized into the following paradigms.

\begin{figure}[htbp]
	\centering
	\subfigure[LLM as controller]{
		\begin{minipage}[b]{0.47\textwidth}
			\includegraphics[width=1\textwidth]{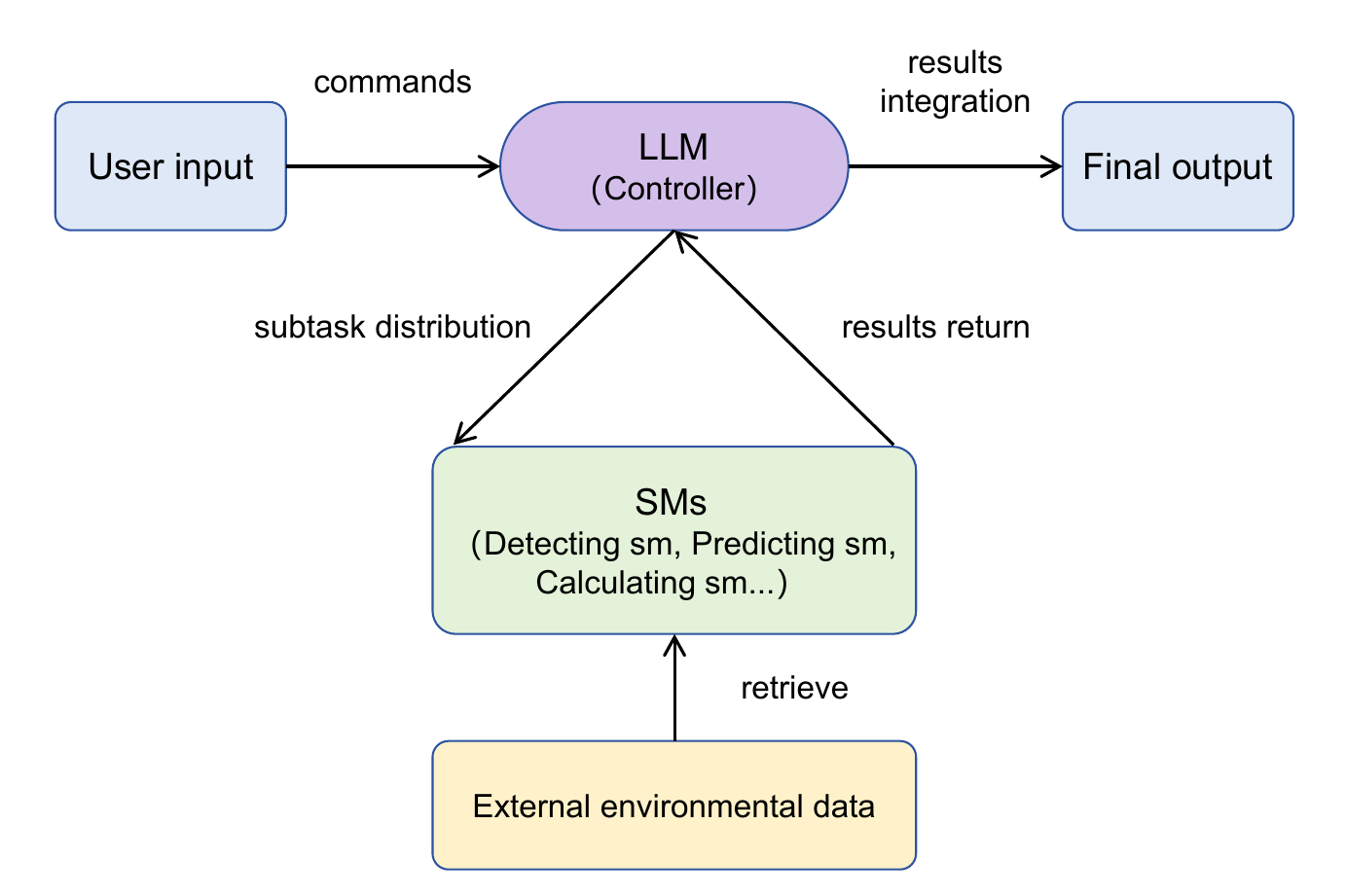}
		\end{minipage}
		\label{Fig1}
	}
	\subfigure[LLM-enhanced SMs]{
		\begin{minipage}[b]{0.47\textwidth}
			\includegraphics[width=1\textwidth]{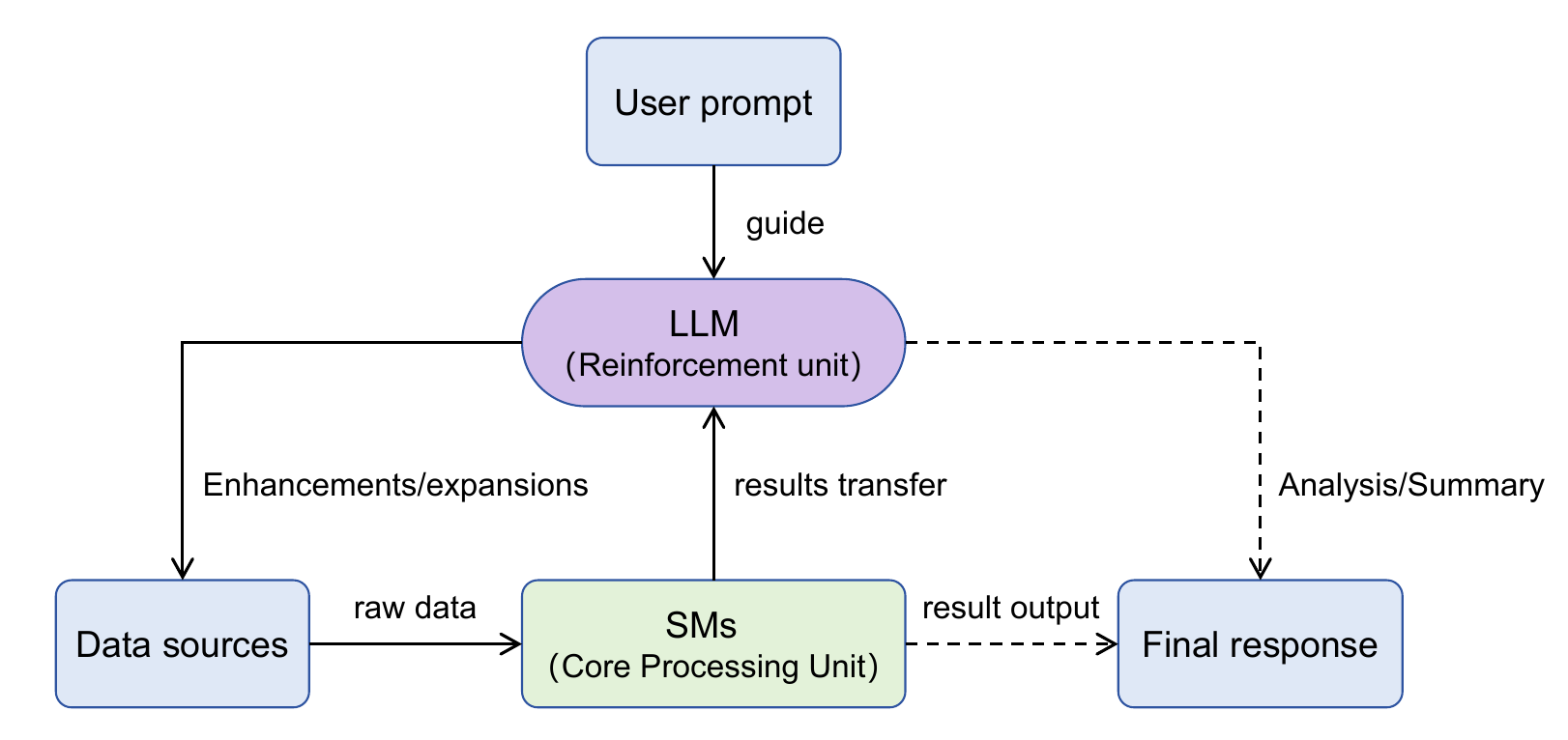}
		\end{minipage}
		\label{Fig2}
	}
	\subfigure[SMs-enhanced LLM]{
		\begin{minipage}[b]{0.47\textwidth}
			\includegraphics[width=1\textwidth]{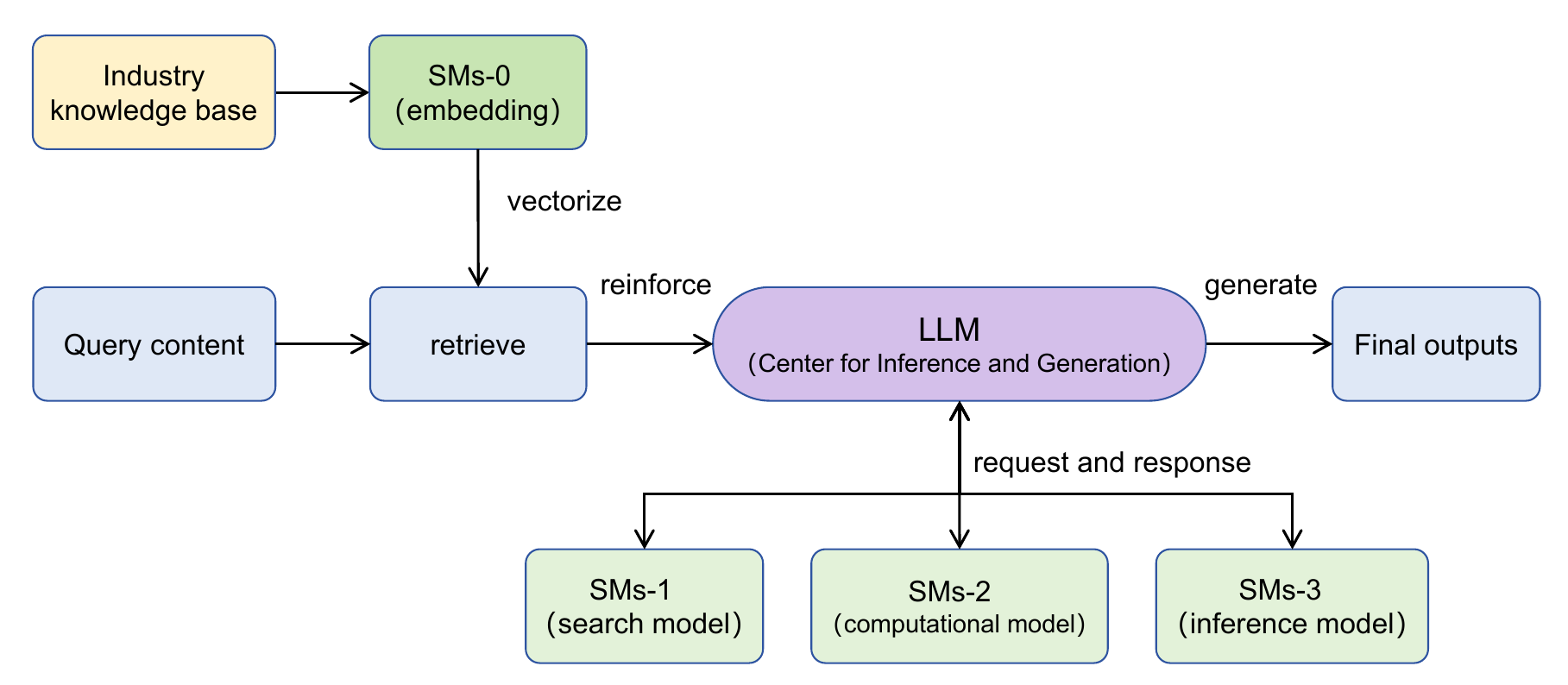}
		\end{minipage}
		\label{Fig3}
	}
	\subfigure[Hybrid collaboration]{
		\begin{minipage}[b]{0.47\textwidth}
			\includegraphics[width=1\textwidth]{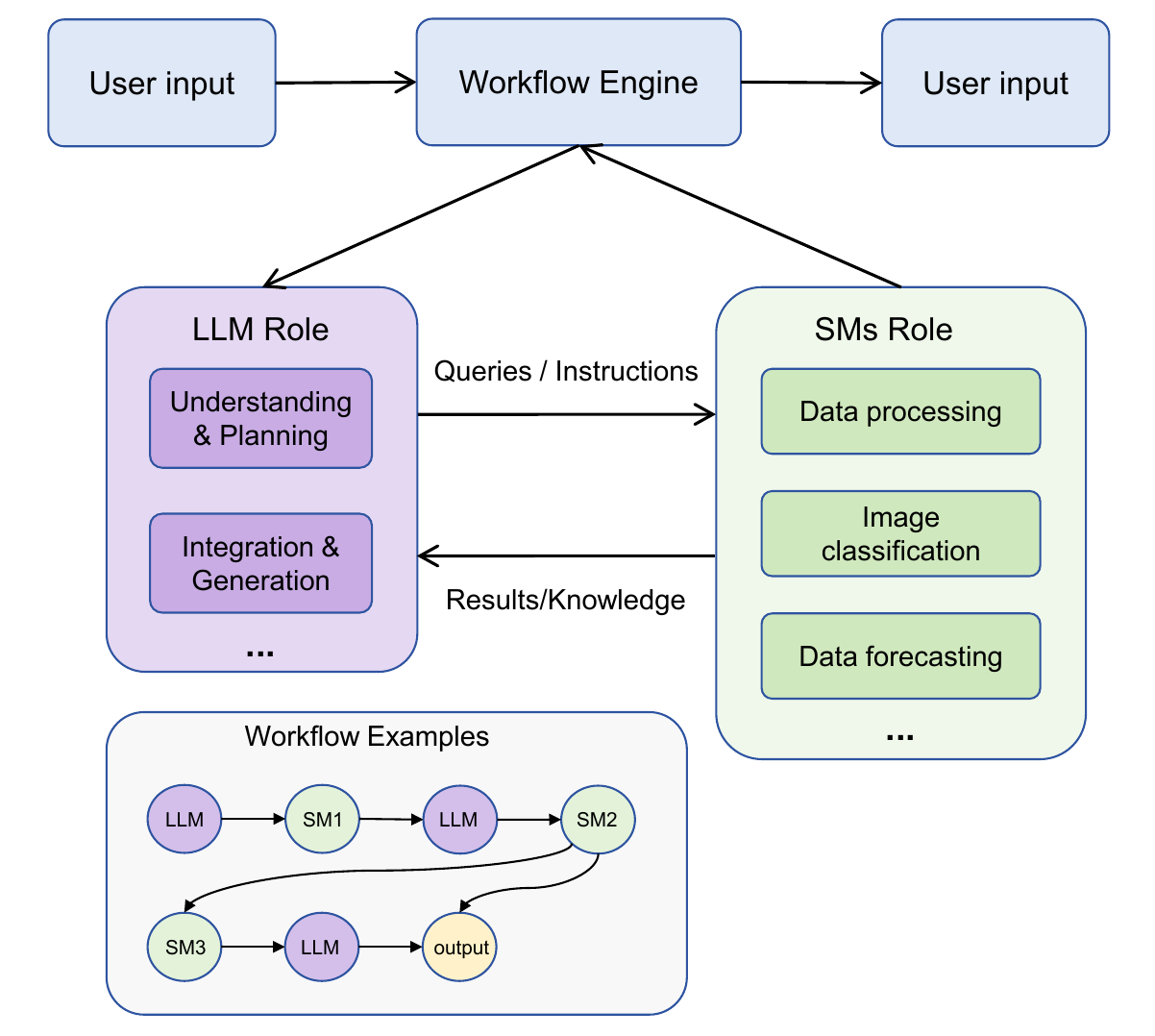}
		\end{minipage}
		\label{Fig4}
	}
	\caption{\color{black}Different paradigms of LLM and SM collaboration.}
	\label{fig:all_paradigms}
\end{figure}

\begin{itemize}
\item \textit{LLM as Controller}: In this paradigm, the LLM receives user instructions or perceives environmental information, decomposes and plans tasks, and invokes external tools (including various SMs) to execute specific subtasks. The LLM is responsible for understanding intentions, integrating information, and generating the final results. Typical frameworks include Hugging-GPT \cite{shen2023hugginggpt}, AutoGPT \cite{wu2023autogen}, etc., which provide mechanisms such as tool invocation, memory management, and agent construction, as shown in Fig. \ref{Fig1}.

\item \textit{LLM-Enhanced SMs}: In this paradigm, using LLMs to enhance the performance or ease of use of SMs is focused on. For example, using LLMs to generate the data required for the training of SMs, explaining the prediction results of SMs, or converting the outputs of SMs into natural language reports, as shown in Fig. \ref{Fig2}.

\item \textit{SMs-Enhanced LLMs}: In this paradigm, SMs is utilized to provide LLMs with precise domain knowledge, factual basis, or computational capabilities, so as to overcome issues such as insufficient professional domain knowledge, inaccurate calculations, or the generation of hallucinations in LLMs. Retrieval-Augmented Generation (RAG) is a typical example of this paradigm. By retrieving relevant domain knowledge bases, which can be processed or constructed by SMs, SMs assist LLMs in generating more accurate and reliable content, as shown in Fig. \ref{Fig3}.

\item \textit{Hybrid Collaboration}: Combining the characteristics of the above paradigms, dynamically adjusting the roles and interaction methods of LLMs and SMs according to the task requirements to form a more complex collaborative workflow. For example, the LLM conducts a preliminary diagnosis, invokes SMs for data verification and in-depth analysis, and then the LLM integrates the information to generate the final report and maintenance suggestions. Frameworks such as LlamaIndex, LangGraph, and MetaGPT provide the ability to build complex workflows, as shown in Fig. \ref{Fig4}. 
\end{itemize}

\begin{table*}[htbp]
	\centering
	\caption{Comparison of Various Collaboration Paradigms between LLMs and SMs}
	\setlength{\tabcolsep}{1.5mm}{
		\begin{tabular}{p{3.5cm}p{3cm}p{4.2cm}p{4.3cm}}
			\specialrule{0.1em}{1pt}{1pt}
			\specialrule{0.1em}{1pt}{3pt}
			\textbf{Paradigm} & \textbf{Advantages} & \textbf{Disadvantages} & \textbf{Applicable Scenarios} \\
			\midrule
			LLM as controller & Strong planning and integration capability; high flexibility & Relies on LLM stability, prone to errors; inefficient planning & Autonomous planning and multi-step task execution; exploratory tasks and open-ended problem solving \\
			LLM enhancing SMs & Addresses limitations in SMs’ data interpretation or explanation capability & Auxiliary information may be inaccurate; relies on the quality of LLM-generated content & Data annotation and generation; natural language interpretation and report generation for SM predictions \\
			SMs enhancing LLM & Improves LLM accuracy and reduces hallucination & SM performance and knowledge base quality affect enhancement effect & Private knowledge-based Q\&A (RAG); high-precision computing; domain-specific applications \\
			Hybrid collaboration mode & High adaptability, flexible policy adjustment based on task requirements & Complex design and coordination; increased debugging and maintenance cost & Complex collaborative tasks; multi-step workflows; comprehensive applications \\
			\specialrule{0.1em}{3pt}{1pt}
			\specialrule{0.1em}{1pt}{1pt}
	\end{tabular}}
	\label{tab:paradigm_comparison}
\end{table*}

The comparison of advantages, disadvantages, and applicable scenarios of different paradigms are listed in Table~\ref{tab:paradigm_comparison}.
There are already some collaboration frameworks for large and small models in the industry, such as AutoGPT, LangChain, LlamaIndex, AutoGen, etc. These mainstream frameworks provide powerful basic capabilities, meanwhile, there are still some deficiencies when they are directly applied to the intelligent maintenance scenarios in industrial fields:

\begin{itemize}

\item \textit{Domain adaptability:} General-purpose frameworks often lack in-depth optimization support for the specific data types in the industrial field, such as high-frequency time-series data, sensor array data, and industrial control system logs.
\item \textit{Real-time performance and reliability:} Industrial scenarios have extremely high requirements for the real-time performance of fault diagnosis and the reliability of operation and maintenance decisions. The invocation delay of LLMs in existing frameworks, potential hallucination problems, and the stability of complex workflows may be difficult to meet these requirements.

\item \textit{Integration complexity:} Integrating various SMs widely deployed in existing industrial systems, such as fault prediction models, root cause analysis models, and optimal scheduling algorithms, into general-purpose frameworks often requires a large amount of customized development and interface adaptation work.

\item \textit{Knowledge representation and integration:} How to effectively integrate the mechanism knowledge and expert experience in the industrial field with real-time operation data, and connect LLMs and SMs through a unified intermediate representation (such as knowledge graphs and semantic layers) is a challenge that existing frameworks have not fully solved.

\end{itemize}

\section{Proposed Sentosa\_LLM Framework}

\begin{figure*}[htbp]
	\centering
	\includegraphics[width=0.95\textwidth]{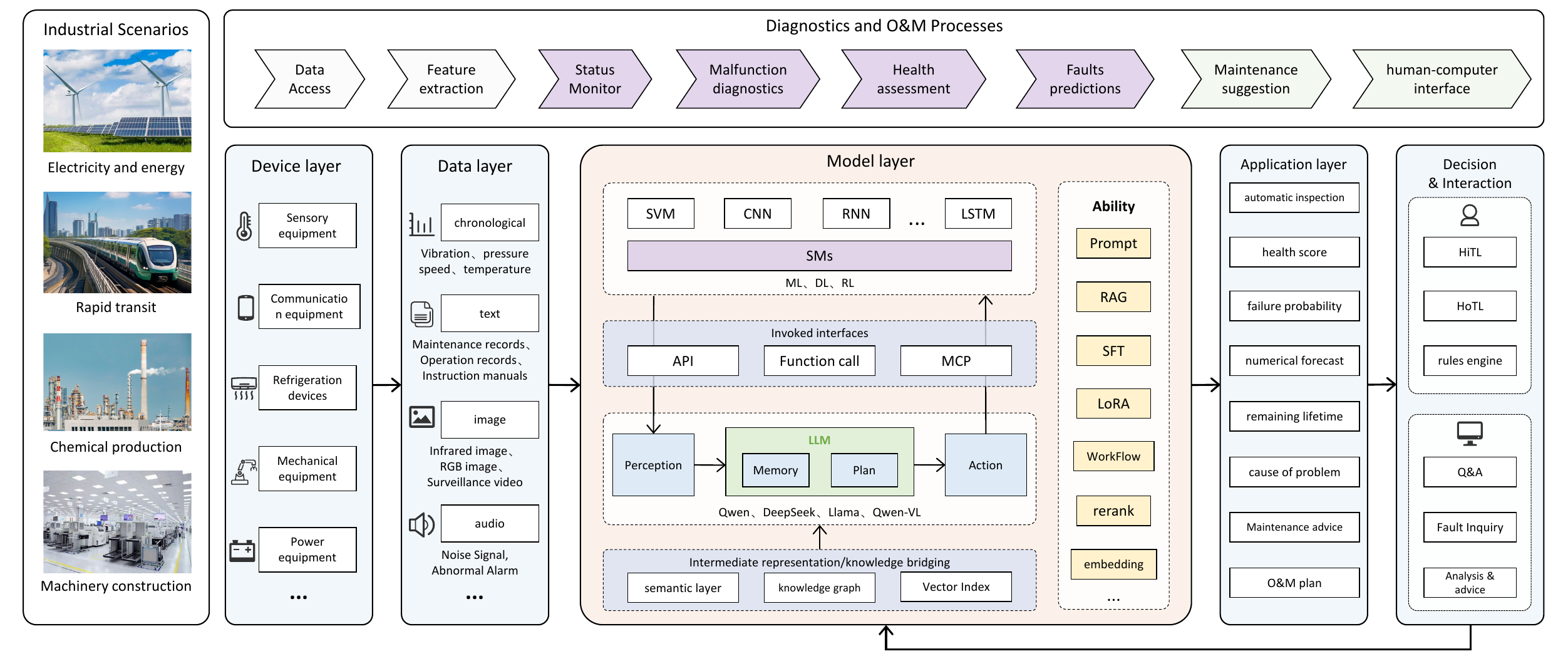}
	\caption{Architecture of the Sentosa\_LLM framework for collaboration between large and small models.}
	\label{Fig5}
\end{figure*}

To overcome the limitations of existing frameworks in industrial applications and deeply integrate the intelligent  maintenance processes, a technical framework for the fusion and collaboration of large and small models for intelligent maintenance in industrial scenarios is proposed, named Sentosa\_LLM, which adopts a hierarchical decoupling design, aiming to organically combine the precise computational capabilities of domain-specific SMs with the cognitive reasoning, knowledge integration, and interaction capabilities of LLMs, so as to achieve an efficient, reliable, and easily integrated intelligent solution. The framework is shown in Fig. \ref{Fig5}.

The core of the Sentosa\_LLM framework is composed of the Equipment Layer, Data Layer, Model Layer, Application Layer, and Decision \& Interaction Layer. 
Model Layer is the core of the Sentosa\_LLM framework, responsible for data processing, analysis, modeling, and reasoning, reflecting the deep integration of large and small models. This layer mainly contains two sub-modules that work collaboratively: Domain-specific SMs module and LLM Collaboration module. 
The Sentosa\_LLM framework uses the following key technologies to achieve interaction and integration with SMs and external knowledge:

\begin{enumerate}

\item Intermediate Representation and Knowledge Bridging, a unified knowledge representation constructed through the Industrial Knowledge Graph, Semantic Layer, and Vector Database to connect structured data, unstructured text, domain knowledge, and model outputs. 

\item Prompt Engineering, carefully designing and optimizing the instructions input to the LLM to guide it in task understanding, planning, and execution. 

\item  RAG, use the vector database and knowledge graph to retrieve relevant context information to enhance the accuracy and reliability of the content generated by the LLM. 

\item  Workflow Engine, responsible for orchestrating and scheduling the collaborative workflow between the LLM and SMs based on predefined templates or dynamically, and automatically executing a series of complex task flows that include data processing, model invocation, information integration, and result generation. 

\item  Tool Invocation, defining standardized interfaces (such as APIs, Function Calling, Model Context Protocol) to enable the LLM to call the models in the SMs module as tools to execute professional computational tasks.
\end{enumerate}

Through the above hierarchical design and collaborative mechanism, the Sentosa\_LLM framework can effectively integrate multi-source heterogeneous data, combine the advantages of large and small models, run through the complete process of intelligent maintenance, and ultimately achieve more accurate, intelligent, and efficient intelligent maintenance in industrial scenarios.

Furthermore, in this framework, the interpretability of model results is achieved through "transparent model design + traceable interaction processes + deep integration of domain knowledge + multi-dimensional evaluation and verification." Specifically, the measures are as follows:

\begin{enumerate}
\item[\textbullet] The SMs is responsible for executing interpretable rule-based reasoning (such as fault threshold judgment based on physical formulas).

\item[\textbullet] The LMs provides natural language explanations and links industry knowledge (e.g., "Based on historical cases, this temperature anomaly may be caused by aging of the component").

\item[\textbullet] The SMs and the LMs form a complete interpretable chain through a logging system and visualization tools, ultimately enhancing engineers' trust in maintenance recommendations and execution efficiency.

\end{enumerate}

\section{Case Study I: Energy Saving of the Equipment Rooms of Communication Operators}

\subsection{Background}

With the popularization of 5G technology and the expansion of data centers, the operation and maintenance challenges of communication operators’ equipment rooms have intensified. The equipment in the rooms features high energy consumption and complex faults. The aim of the experiment is to verify the practical effects of the Sentosa\_LLM framework, through the collaboration of large and small models, in improving the accuracy of temperature prediction of equipment room, optimizing air conditioning control, and saving energy and reducing power consumption.

\subsection{Datasets}

A dataset collected from an equipment room operated by China Unicom was utilized, covering the period from January 1, 2024, to December 31, 2024. The dataset contains multi-source heterogeneous data that capture both internal operational conditions and external environmental variables. Specifically, the time-series data include temperature and humidity readings from multiple indoor sensors, operational status logs of the air conditioning system, and measurements from electrical meters, all recorded at two-minute intervals. External environmental data, obtained hourly through external tool invocation by a large-scale model, consist of outdoor temperature, humidity, and extreme weather indicators. Additionally, unstructured textual data are included, comprising operation and maintenance logs along with the air conditioning equipment manual.

\subsection{Results}

The baseline model is a long short-term memory (LSTM) network. In parallel, an energy-saving control strategy is implemented using a rule-based expert system developed with a predefined rule engine. A comparative analysis is conducted across three configurations: the LSTM model trained with raw data, the LSTM model trained with data augmented by a LLM, and LSTM outputs further refined through LLM-based correction that incorporates high-order semantic features. Results are reported in Table~\ref{tab:temp_prediction_comparison}.

The findings demonstrate that under the Sentosa\_LLM framework, the prediction performance of the LSTM in temperature forecasting for the equipment room is substantially improved through collaborative enhancement. Specifically, by leveraging LLM-driven analysis of historical records and domain-specific textual information to generate targeted augmented data, the mean absolute error (MAE) and root mean square error (RMSE) are reduced by 15.4\% and 17.5\%, respectively. Moreover, through the incorporation of contextual signals derived from both external environmental conditions and internal system states, the refined predictions yield additional reductions in MAE, RMSE, and mean absolute percentage error (MAPE) of approximately 18\% compared with the model trained on augmented data alone.
Table~\ref{tab:temp_energy_comparison} further presents a comparative evaluation of temperature control performance and the corresponding energy-saving outcomes over one-week intervals. The evaluation is based on real-world deployments conducted within the same equipment room under different prediction schemes and control strategies.

\begin{table}[htbp]
	\centering
	\caption{Comparison of Temperature Prediction Results under Different Modes}
	\setlength{\tabcolsep}{1mm}{
		\begin{tabular}{cccccc}
			\specialrule{0.1em}{1pt}{1pt}
			\specialrule{0.1em}{1pt}{3pt}
			Mode & Item & MAE & RMSE & Avg MAE & Avg RMSE \\
			\midrule
			\multirow{3}[1]{*}{\makecell{LSTM \\ (Only)}} & 1 & 0.45 & 0.62 & \multirow{3}[1]{*}{0.453} & \multirow{3}[1]{*}{0.61} \\
			& 2 & 0.44 & 0.60 & & \\
			& 3 & 0.47 & 0.61 & & \\
						\midrule
			\multirow{3}[0]{*}{\makecell{LSTM \\ (LLM data augment)}} & 1 & 0.38 & 0.51 & \multirow{3}[0]{*}{0.383} & \multirow{3}[0]{*}{0.503} \\
			& 2 & 0.40 & 0.52 & & \\
			& 3 & 0.37 & 0.48 & & \\
						\midrule
			\multirow{3}[1]{*}{\makecell{LSTM + \\ LLM Correct}} & 1 & 0.31 & 0.42 & \multirow{3}[1]{*}{0.300} & \multirow{3}[1]{*}{0.403} \\
			& 2 & 0.30 & 0.40 & & \\
			& 3 & 0.29 & 0.39 & & \\
			\specialrule{0.1em}{3pt}{1pt}
			\specialrule{0.1em}{1pt}{1pt}
	\end{tabular}}
	\label{tab:temp_prediction_comparison}%
\end{table}

\begin{table}[htbp]
	\centering
	\caption{Comparison of Temperature Control Accuracy and Energy Consumption under Different Strategies}
	\setlength{\tabcolsep}{2.4mm}{
		\begin{tabular}{lccc}
			\specialrule{0.1em}{1pt}{1pt}
			\specialrule{0.1em}{1pt}{3pt}
			\multirow{2}{*}{Strategy} &
			\multirow{2}{*}{Mode} &
			Accuracy&Total energy \\
			& & (24--28$^\circ$C) & (kWh) \\
			\midrule
			Default & N/A & 86.5\% & 15837 \\
			Rule-based & LSTM Only & 94.5\% & 14214 \\
			LLM & SM+LLM Correct & 97.8\% & 12525 \\
			\specialrule{0.1em}{3pt}{1pt}
			\specialrule{0.1em}{1pt}{1pt}
	\end{tabular}}
	\label{tab:temp_energy_comparison}
\end{table}

\section{Case Study II: Fault Detection of Energy Storage Power Station}

\subsection{Background}
Energy storage power stations play a crucial role in balancing power grid supply and demand and improving energy utilization efficiency. However, energy storage power stations face safety hazards such as tiny internal short circuits and thermal runaway in batteries, as well as deficiencies in battery health and lifespan assessment. These issues and potential risks threaten the safe operation of energy storage power stations. Therefore, it is urgent to adopt advanced intelligent operation and maintenance technologies to guarantee their safe, stable, and efficient operation.
This experiment takes the early diagnosis of tiny internal short-circuit faults in energy storage power stations as an example to study the comparison between deep learning models and the collaboration of large and small models.

\subsection{Dataset}

We utilize a dataset collected from one of energy storage power stations with capacity of 511MW/1071.5MWh, which is located in Shandong Province. The dataset is derived from electrochemical time-series data collected at the single-cell level using a Battery Management System (BMS), encompassing voltage, current, and temperature sampled with one-second intervals and 2-years duration.  

\subsection{Results}


\begin{table*}[htbp]
	\centering
	\caption{Analysis of Representative Fault Detection Case}
	\setlength{\tabcolsep}{4mm}{
		\begin{tabular}{ll}
			\specialrule{0.1em}{1pt}{1pt}
			\specialrule{0.1em}{1pt}{3pt}
			\textbf{Item} & \textbf{Description} \\

			\specialrule{0.1em}{1pt}{3pt}
			Location & Energy storage power station in Shandong Province \\
			Fault Module & \#B23 \\
			Detected Fault & Micro internal short circuit ($\mu$ISC), estimated resistance: 0.6~$\Omega$ \\
			Initial Alert & Issued by the small model on Day 9; capacity difference standard deviation reached 0.63\% \\
			Fault Confirmation & Large model identified cell \#7 and predicted the fault development trajectory \\
			Physical Verification & Disassembly revealed a 3mm micro-perforation in the separator \\
			Mean Time to Repair (MTTR) & Reduced by 41\% \\
			Operation and Maintenance (O\&M) Cost & Decreased by 27\% \\
			Validation Basis & Field-tested in a 500MW/1000MWh energy storage station (Shandong Province) \\
			\specialrule{0.1em}{3pt}{1pt}
			\specialrule{0.1em}{1pt}{1pt}
	\end{tabular}}
	\label{Table1}
\end{table*}

We still consider to use the LSTM network as the baseline small model. Initially, the data are denoised using Kalman filtering. Feature extraction is then performed through wavelet transform and empirical mode decomposition (EMD), after which the LSTM network is trained to detect minor internal short-circuit faults.

Within the Sentosa\_LLM framework, users query the health status of battery arrays through LMs. Based on integrated knowledge bases and system topologies, LMs invoke small LSTM models to generate diagnostic results. These results are subsequently subjected to spatio-temporal consistency checks, such as determining whether abnormal clusters are located within the same cooling branch or whether a voltage drop precedes a temperature increase. This process enables the correction of diagnostic outputs produced by the small models, thereby enhancing detection accuracy. Maintenance recommendations are then generated and delivered by the LMs.



The proposed framework significantly improves early warning for minor internal short circuits in lithium battery systems by extending detection lead time while maintaining high diagnostic accuracy. It also achieves low reasoning latency and memory usage, supporting real-time industrial deployment. For each potential risk, the system provides actionable recommendations, enabling proactive and intelligent fault management in complex environments.

\section{Conclusion}
We investigate the application of large models in intelligent maintenance for industrial systems. Through a systematic review of existing methods and implementations, we identify their strengths and limitations. To address these challenges, we propose a unified framework that integrates the complementary strengths of large and small models.

The framework adopts a hierarchical architecture to enable the fusion of multi-source heterogeneous data. We combine the semantic reasoning capabilities of large models with the low-latency and resource-efficient features of small models to support end-to-end decision-making across the maintenance lifecycle.
We validate the proposed approach in two representative industrial scenarios, including telecommunications equipment rooms and energy storage systems. The results demonstrate its effectiveness, scalability, and potential to serve as a foundational paradigm for future intelligent maintenance solutions.

\bibliographystyle{ieeetr}
\bibliography{LMM}

\end{document}